\documentclass[12pt,epsf]{article}
\usepackage[xdvi]{graphicx}
\newcommand{\beq}{\begin{equation}}
\newcommand{\eeq}{\end{equation}}
\newcommand{\bea}{\begin{eqnarray}}
\newcommand{\eea}{\end{eqnarray}}

\def\pe2{p_E^2}

\makeatletter

\@addtoreset{equation}{section}
\makeatother

\begin{document}
\newcommand{\mpl}{M_{\mathrm{Pl}}}
\setlength{\baselineskip}{18pt}
\begin{titlepage}
\begin{flushright}
KOBE-TH-07-01 
\end{flushright}

\vspace{1.0cm}
\begin{center}
{\Large\bf Calculable One-Loop Contributions \\
\vspace*{5mm}
to S and T Parameters   
in the Gauge-Higgs \\
\vspace*{5mm}
Unification} 
\end{center}
\vspace{25mm}

\begin{center}
{\large
C. S. Lim
\footnote{e-mail : lim@kobe-u.ac.jp}
and Nobuhito Maru\footnote{e-mail : 
maru@people.kobe-u.ac.jp}}
\end{center}
\vspace{1cm}
\centerline{{\it Department of Physics, Kobe University,
Kobe 657-8501, Japan}}
%
%
\vspace{2cm}
\centerline{\large\bf Abstract}
\vspace{0.5cm}
We investigate the one-loop contributions to 
S and T oblique parameters in gauge-Higgs unification. 
We show that these parameters are finite in five dimensional space-time, 
but are divergent in more than five dimensions. 
Remarkably, however, we find that a particular linear 
combination of S and T parameters, $S - 4 \cos \theta_{W} \ T$, becomes finite 
for six dimensional space-time, though each of these 
parameters are divergent. This is because, in the Gauge-Higgs unification 
scenario, the operators relevant for S and T parameters are not independent, 
but are included in a unique higher dimensional gauge invariant operator. 
Thus the predictable linear combination is model independent, irrespectively  
of the detail of the matter content.  
 
\end{titlepage}

%
%
\section{Introduction} 
Solving the hierarchy problem motivates us to go beyond the Standard Model (SM). 
Gauge-Higgs unification is one of the attractive approaches 
to solve the hierarchy problem without supersymmetry. 
In this scenario, 
Higgs field is identified with the zero mode of 
the extra component of the gauge field in higher dimensional gauge theories 
\cite{Manton, Fairlie} 
and the gauge symmetry breaking occurs dynamically 
through the vacuum expectation value of Wilson line phase: the Hosotani mechanism \cite{Hosotani}. 
One of the remarkable features is that the quantum correction 
to the  Higgs mass-squared becomes finite 
thanks to the higher dimensional local gauge invariance, 
once all Kaluza-Klein (K-K) modes are summed up 
in the intermediate state of the loop diagram, 
thus solving the hierarchy problem at quantum level 
(the problem of ``quadratic divergence") \cite{HIL}-\cite{CNP}.

Recently, the scenario has been further developed and extended. In the case of gauge-Higgs unification 
on a simply-connected curved space S$^{2}$ as the extra space, 
the quantum correction to the Higgs mass turns out not only to be finite 
but also to vanish identically \cite{LMH}. 
A similar mechanism is found to work 
in the ``gravity-gauge-Higgs unification" scenario, 
where Higgs field is identified with the extra-space component of 
a higher dimensional graviton field \cite{HLM}. 
The argument of the finiteness can be extended to the two-loop level \cite{MY},  
and for Abelian gauge theories the finiteness is proved to hold 
at any order of the perturbative expansion \cite{Hosotani2}. 
Furthermore, attempts to construct a realistic model 
embodied with the idea of gauge-Higgs unification 
and investigations into the possible applications of the scenario 
in various aspects have been carried out \cite{KLY}-\cite{CPSW}. 

In order to understand the gauge-Higgs unification scenario more deeply 
and to make a realistic model, 
it is important to ask whether there are other finite (``calculable") physical 
observables besides Higgs mass, which are genuine predictions of the scenario.  
Let us note that the reason why the Higgs mass is calculable 
in higher dimensional gauge theories, which are argued to be non-renormalizable, 
is that the gauge-Higgs sector is controlled 
by the higher dimensional local gauge invariance, and 
no local gauge invariant operator responsible for the Higgs mass exists. 
It will be natural to ask whether there exist other calculable observables 
protected by higher dimensional gauge symmetry. 

In this paper we consider the ``oblique" parameters S and T \cite{PT} 
in the framework of Gauge-Higgs unification scenario, 
as one of the good candidates of such calculable physical observables. 
The parameters are defined as   
\bea
S &=& -\frac{16 \pi}{g^{2} \tan \theta_{W}} \Pi'_{{\rm 3Y}},  
\label{Sdef}\\
T &=& - \frac{4\pi}{g^{2} \sin^2 \theta_W} \frac{\Delta M^{2}}{M_{W}^{2}},  
\label{Tdef}
\eea
where $\Pi'_{3Y} \equiv \frac{d^2}{dp^2}\Pi_{3Y}(p^2)|_{p^{2}=0}$, with $\Pi_{3Y}(p^2)$ being the $g_{\mu\nu}$ part of the gauge boson self-energy between $W^{3}_{\mu}$ and $B_{\mu}$ ($W^{3}_{\mu}, \ B_{\mu}: SU(2)_{L}, U(1)_{Y} \mbox{gauge bosons}$) and $\theta_W$ denotes the  Weinberg angle. $\Delta M^{2} \equiv \delta M_{W^{3}}^{2} - \delta M_{W^{\pm}}^{2}$,  with 
$\delta M_{W^{3}}^{2}, \delta M_{W^{\pm}}^{2}$ being quantum corrections to the neutral and charged 
gauge boson mass-squared.  

The reason to take these parameters as the candidates is twofold. 
First, both of S and T parameters are described 
in four dimensional (4D) space-time by (the coefficients of) 
``irrelevant" gauge invariant operators with higher ($d = 6$) mass dimension, 
such as $(\phi^{\dagger} D_{\mu} \phi)(\phi^{\dagger} D^{\mu} \phi)$ 
for T and $(\phi^{\dagger} W^{a}_{\mu \nu}\frac{\tau^{a}}{2} \phi) B^{\mu \nu}$ 
for S ($\phi$: Higgs doublet, $W^{a}_{\mu \nu}, \ B_{\mu \nu}$: 
field strengths of SU(2)$_{L}$ and U(1)$_{Y}$ gauge fields). 
Since in the gauge-Higgs unification scenario the Higgs $\phi$ is unified with the gauge field, 
there is a possibility that these two operators are also unified 
in a single gauge invariant local operator 
with respect to higher dimensional gauge field, 
whose mass dimension is 6 from 4D point of view. 
This means that the structure of the divergence of S and T parameters 
are not independent 
and some particular linear combination of these parameters 
is expected to be free from UV divergence, 
thus making it theoretically predictable. 
Second the S and T parameters have been severely constrained 
through the precision electro-weak measurements. 
Thus studying these parameters is very useful 
in searching for a viable model based on the scenario.       
  
The model we take in this paper is minimal $SU(3)$ gauge-Higgs unification model 
compactified on an orbifold $S^1/Z_2$ with a triplet fermion as the matter field.  
We confirm by explicit one-loop calculations that our expectation stated above 
is the case.

The S and T parameters are calculated in two approaches. 
One approach is to perform the dimensional regularization 
for the 4D momentum integral before taking the K-K mode sum, 
which has the advantage that the 4D gauge invariance is manifest in each K-K mode, 
though the whole structure of divergence becomes clear only after the mode sum. 
The other one is to take the mode sum before the momentum integral, 
which is also useful to extract the whole structure of possible UV divergence 
and to make the higher dimensional gauge invariance manifest. 
In order to generalize our argument to more than 5D space-times, 
we first derive general formulas where the dimensionality $D$ 
in the dimensional regularization is left arbitrary. 
  
As the result, we show that one-loop contributions to S and T parameters 
are both finite in 5D space-time, 
but are divergent for higher space-time dimensions. 
The remarkable result is that in 6D space-time, 
although one-loop corrections to S and T parameters themselves are certainly divergent, a particular linear combination of these parameters, $S - 4 \cos \theta_{W} \ T$, becomes finite (calculable) and predictable.  
We also show that the ratio of the coefficients in the linear combination 
just coincides with what we obtain from a single gauge invariant operator 
with respect to higher dimensional gauge fields $(D_{L} F_{MN})(D^{L}F^{MN})$, 
which means that the predictable observable is fixed in a model independent way, 
irrespectively of the detail of each model's matter content. 
This is a crucial difference from the situation 
in the universal extra dimension (UED) scenario \cite{UED}. 

This paper is organized as follows. 
In the next section, 
we introduce our model. 
In section 3, the one-loop contribution to the T-parameter in 5D  
is calculated in two different approaches stated above. 
A similar calculation on the one-loop contribution to the S-parameter in 5D 
is given in section 4. 
In section 5, we generalize our results to higher space-time dimensions, 
and discuss the finiteness of a particular linear combination of 
S and T parameters for the 6D case. 
Section 6 is devoted to the summary and some concluding remarks. 
In appendix A, some technical detail of the calculation 
for the T parameter is given.  

\section{The Model}
In this paper, 
we adopt a minimal model of SU(3) gauge-Higgs unification 
with an orbifold $S^1/Z_2$ as the extra space, in order to 
avoid unnecessary complications in investigating 
the divergence structure of the one-loop contributions to 
the S and T parameters, though the predicted Weinberg angle 
is unrealistic, $\sin^{2} \theta_{W} = \frac{3}{4}$.\footnote{There is 
no tree level contribution to these parameters in our model. 
As pointed out in \cite{CCP}, 
if we add an extra $U(1)$ to obtain the desirable Weinberg angle, 
a tree level contribution to the T-parameter appears, 
and a certain constraint on the compactification scale must be imposed.} 
As the matter field we introduce an SU(3) triplet fermion, which we identify 
with ``top and bottom" quarks and their K-K ``excited states", 
although the top quark mass vanishes and 
the bottom quark mass $m_{b} = M_{W}$ in this toy model. 
(For instance, the T parameter is sensitive to the mass splitting between $m_t$ and $m_{b}$, 
not their absolute values, anyway.) 
In this work we neglect the presence of generations. 

The SU(3) symmetry is broken into SU(2) $\times$ U(1) 
by the orbifolding $S^1/Z_2$ and adopting a non-trivial $Z_{2}$ parity assignment 
for the members of an irreducible repr. of SU(3), as stated below. 
The remaining gauge symmetry SU(2) $\times$ U(1) is supposed to be broken 
by the VEV of the K-K zero-mode of $A_{5}$, 
the extra space component of the gauge field behaving as the Higgs doublet, 
through the Hosotani-mechanism \cite{Hosotani}, 
though we do not address the question how the VEV is obtained 
by minimizing the loop-induced effective potential for $A_{5}$ \cite{KLY}.      

The lagrangian is simply given by 
\bea
{\cal L} = -\frac{1}{2} \mbox{Tr}  (F_{MN}F^{MN}) 
+ i\bar{\Psi}D\!\!\!\!/ \Psi
\label{lagrangian}
\eea
where $\Gamma^M=(\gamma^\mu, i \gamma^5)$, 
\bea
F_{MN} &=& \partial_M A_N - \partial_N A_M -i g_{5} [A_M, A_N]~(M,N = 0,1,2,3,5), \\
D\!\!\!\!/ &=& \Gamma^M (\partial_M -ig_{5} A_M) \ \ 
(A_{M} = A_{M}^{a} \frac{\lambda^{a}}{2} \ 
(\lambda^{a}: \mbox{Gell-Mann matrices})),  \\
\Psi &=& (\psi_1, \psi_2, \psi_3)^T.
\eea
The periodic boundary conditions are imposed along $S^1$ for all fields and 
the non-trivial $Z_2$ parities are assigned for each field  as follows, 
\bea 
\label{z2parity} 
A_\mu = 
\left(
\begin{array}{ccc}
(+,+) & (+,+) & (-,-) \\
(+,+) & (+,+) & (-,-) \\
(-,-) & (-,-) & (+,+) 
\end{array}
\right), \ \ 
A_5 = 
\left(
\begin{array}{ccc}
(-,-) & (-,-) & (+,+) \\
(-,-) & (-,-) & (+,+) \\
(+,+) &(+,+) & (-,-)
\end{array}
\right), 
\eea
\bea 
\label{fermizero} 
\Psi = 
\left(
\begin{array}{cc}
\psi_{1L}(+,+) + \psi_{1R}(-, -) \\
\psi_{2L}(+,+) + \psi_{2R}(-, -) \\
\psi_{3L}(-,-) + \psi_{3R}(+, +) \\
\end{array}
\right),
\eea
where $(+,+)$ means that $Z_2$ parities are even at the fixed points $y=0$ 
and $y = \pi R$, for instance. $y$ is the fifth coordinate and 
$R$ is the compactification radius. 
$\psi_{1L} \equiv \frac{1}{2}(1-\gamma_5)\Psi$, etc. 
A remarkable feature of this manipulation of ``orbifolding" is that 
in the gauge-Higgs sector, 
exactly what we need for the formation of the standard model is obtained at low energies; 
one can see that $SU(3)$ is broken to $SU(2)_{L} \times U(1)_{Y}$ and the Higgs doublet 
$\phi = (\phi^{+}, \phi^{0})^{t}$ emerges. 
Namely the K-K zero-mode of the gauge-Higgs sector takes the form, 
\bea
A^{(0)}_{\mu} = \frac{1}{2}
\left(
\begin{array}{ccc}
W^{3}_{\mu}+ \frac{B_{\mu}}{\sqrt{3}} & \sqrt{2} W^{+}_{\mu} & 0 \\
\sqrt{2} W^{-}_{\mu} & - W^{3}_{\mu}+ \frac{B_{\mu}}{\sqrt{3}} & 0 \\
0& 0 & -\frac{2}{\sqrt{3}} B_{\mu}   
\end{array}
\right) , \ \ 
A_5^{(0)} = \frac{1}{\sqrt{2}}
\left(
\begin{array}{ccc}
0 & 0 & \phi^{+} \\
0 & 0 & \phi^{0} \\
\phi^{-} & \phi^{0\ast} & 0 
\end{array}
\right),   
\eea
with $W_\mu^{3}, \ W_{\mu}^{\pm}$, $B_\mu$ being the $SU(2)_{L}, U(1)_{Y}$ gauge fields, 
respectively, while in the zero-mode of the triplet fermion $t_{R}$ is lacking, 
\bea
\Psi^{(0)} = 
\left(
\begin{array}{cc}
t_{L} \\
b_{L} \\
b_{R} \\
\end{array}
\right). 
\eea
The VEV to break $SU(2)_{L} \times U(1)_{Y}$ is written as 
\beq 
\langle A_{5} \rangle = 
\frac{v}{2} \ \lambda_{6} \ \ (\langle \phi^{0} \rangle = \frac{v}{\sqrt{2}}). 
\eeq  

Following these boundary conditions, 
K-K mode expansions for the gauge fields 
and the fermions are carried out: 
\bea
A_{\mu,5}^{(+,+)}(x,y) &=& \frac{1}{\sqrt{2 \pi R}} 
\left[
A_{\mu,5}^{(0)}(x) + \sqrt{2} \sum_{n=1}^\infty A_{\mu,5}^{(n)}(x) 
\cos (ny/R)
\right], \\
A_{\mu,5}^{(-,-)}(x,y) &=& \frac{1}{\sqrt{\pi R}} 
\sum_{n=1}^\infty A_{\mu,5}^{(n)}(x) 
\sin (ny/R), \\
\psi_{1L, 2L, 3R}^{(+,+)}(x,y) &=& \frac{1}{\sqrt{2 \pi R}} 
\left[
\psi_{1L, 2L, 3R}^{(0)}(x) 
+ \sqrt{2} \sum_{n=1}^\infty \psi_{1L,2L,3R}^{(n)}(x) \cos (ny/R)
\right], \\
\psi_{3L,1R,2R}^{(-,-)}(x,y) &=& i \ \frac{1}{\sqrt{\pi R}} 
\sum_{n=1}^\infty \psi_{3L,1R,2R}^{(n)}(x) \sin (ny/R). 
\label{KK}
\eea 

In this paper we discuss one-loop contributions to the S and T parameters due to fermions, 
which potentially give sizable effects, 
though, e.g., the contribution due to gauge boson self interactions 
to the T parameter is handled by $U(1)_{Y}$ gauge coupling $g'$ 
and is expected to be relatively not significant.  
For such purpose, 
only the term containing fermions, 
${\cal L}_{{\rm fermion}}  = i\bar{\Psi}D\!\!\!\!/ \Psi$, 
in the lagrangian (\ref{lagrangian}) is enough to consider. 
Substituting the above K-K expansions for the triplet fermion and the zero-modes for the gauge-Higgs bosons in the term 
and integrating over the fifth coordinate $y$,  
we obtain a 4D effective Lagrangian:  
\bea
{\cal L}_{{\rm fermion}}^{(4D)} 
&=& 
\sum_{n=1}^{\infty} \left\{  
i(\bar{\psi}_1^{(n)}, \bar{\psi}_2^{(n)}, \bar{\psi}_3^{(n)}) 
\gamma^\mu \partial_\mu  
\left(
\begin{array}{c}
\psi_1^{(n)} \\
\psi_2^{(n)} \\
\psi_3^{(n)}
\end{array}
\right) \right. \nonumber \\ 
&& \left. 
+\frac{g}{2} (\bar{\psi}_1^{(n)}, \bar{\psi}_2^{(n)}, \bar{\psi}_3^{(n)})  
\left(
\begin{array}{ccc}
W^{3}_{\mu}+ \frac{B_{\mu}}{\sqrt{3}} & \sqrt{2} W^{+}_{\mu} & 0 \\
\sqrt{2} W^{-}_{\mu} & - W^{3}_{\mu}+ \frac{B_{\mu}}{\sqrt{3}} & 0 \\
0& 0 & -\frac{2}{\sqrt{3}} B_{\mu}   
\end{array}
\right) 
\gamma^{\mu} 
\left(
\begin{array}{c}
\psi_1^{(n)} \\
\psi_2^{(n)} \\
\psi_3^{(n)}
\end{array}
\right) \right. \nonumber  \\ 
&& \left. 
- (\bar{\psi}_1^{(n)}, \bar{\psi}_2^{(n)}, \bar{\psi}_3^{(n)})    
\left(
\begin{array}{ccc}
m_{n} & 0 & 0 \\
0 & m_{n} & -m \\
0& -m  & m_{n} 
\end{array}
\right)
\left(
\begin{array}{c} 
\psi_1^{(n)} \\
\psi_2^{(n)} \\
\psi_3^{(n)}
\end{array}
\right) \right\} \nonumber \\ 
&&  + i \bar{t}_{L} \gamma^{\mu} \partial_{\mu} t_{L} 
+  \bar{b} (i \gamma^{\mu} \partial_{\mu} - m) b 
\nonumber \\ 
&& +\frac{g}{\sqrt{2}} (\bar{t}\gamma_{\mu} L b  W^{+\mu} 
+ \bar{b}\gamma_{\mu} L t W^{-\mu}) 
+ \frac{g}{2} (\bar{t}\gamma_{\mu} L t 
- \bar{b}\gamma_{\mu} L b) W_{3}^{\mu} \nonumber \\ 
&& + \frac{\sqrt{3}g}{6} (\bar{t}\gamma_{\mu} L t 
+ \bar{b}\gamma_{\mu} L b -2\bar{b}\gamma_{\mu} R b) B^{\mu}, 
\label{4Deffaction}
\eea
where $L \equiv \frac{1}{2}(1 -\gamma_5)$, $m_{n} = \frac{n}{R}$. 
$g = \frac{g_{5}}{\sqrt{2\pi R}}$ is the 4D gauge coupling 
and $m = \frac{gv}{2} (= M_{W})$ is the bottom quark mass $m_{b}$. 
Let us note that non-zero K-K modes have both chiralities, 
as is seen in (\ref{fermizero}), 
and their gauge interactions are vector-like, 
described by Dirac particles constructed as 
\bea 
\psi_{1,2,3}^{(n)} &=& \psi_{1,2,3 R}^{(n)} + \psi_{1,2,3,L}^{(n)}~(n > 0). 
\eea
Concerning fermion zero-mode, 
$b = b_{R} + b_{L}$ is a Dirac spinor, while $t$ quark remains a Weyl spinor $t_{L}$. 
We realize that the fermion zero-modes have exactly the same gauge interaction 
as those in the SM, though $\sin^{2} \theta_{W} = \frac{3}{4}$ and 
\bea
m_t = 0, \quad m_b = m \ (= M_{W}).  
\label{qmass}
\eea
In deriving the 4D effective Lagrangian (\ref{4Deffaction}), 
a chiral rotation 
\beq  
\psi_{1,2,3} \ \to \ e^{-i\frac{\pi}{4}\gamma_{5}} \psi_{1,2,3}  
\eeq
has been made in order to get rid of $i \gamma_{5}$.

We easily see that the mass matrix for the non-zero K-K modes 
can be diagonalized by use of the mass eigenstates 
$\tilde{\psi}_{2}^{(n)}, \ \tilde{\psi}_{3}^{(n)}$,   
\bea 
\pmatrix{ 
\psi_{1}^{(n)} \cr 
\tilde{\psi}_{2}^{(n)} \cr 
\tilde{\psi}_{3}^{(n)} \cr 
} 
= U 
\pmatrix{ 
\psi_{1}^{(n)} \cr 
\psi_{2}^{(n)} \cr 
\psi_{3}^{(n)} \cr 
}, \ \ \  
U =\frac{1}{\sqrt{2}}
\left(
\begin{array}{ccc}
\sqrt{2} & 0 & 0 \\
0 & 1 & -1 \\
0 & 1 & 1 
\end{array}
\right), 
\eea
as 
\bea 
U \ 
\left(
\begin{array}{ccc}
m_{n} & 0 & 0 \\
0 & m_{n} & -m  \\
0& -m  & m_{n} 
\end{array}
\right) 
\ U^{\dagger} 
= 
\left(
\begin{array}{ccc}
m_n & 0 & 0 \\
0 & m_n + m & 0 \\
0 & 0 & m_n-m 
\end{array}
\right). 
\eea
Note that a mixing occurs between the $SU(2)$ doublet component 
and singlet component, accompanied by the mass splitting $m_{n} \pm m$.  
Each of mass-eigenvalues has a periodicity with respect to $m$: 
$m_{n} \pm (m + \frac{1}{R}) = m_{n \pm 1} \pm m$, 
which is a remarkable feature of gauge-Higgs unification, not shared by 
the UED scenario, where the masses of non-zero K-K modes behave as 
$\sqrt{m_{n}^{2} + m^{2}}$.

In terms of the mass-eigenstates for non-zero K-K modes, 
the lagrangian reads as 
\bea
&&{\cal L}_{{\rm fermion}}^{(4D)}  
= \sum_{n=1}^{\infty} \left\{  (\bar{\psi}_1^{(n)}, 
\bar{\tilde{\psi}}_2^{(n)}, \bar{\tilde{\psi}}_3^{(n)}) 
\right. \nonumber \\ 
&& \left. \times \left(
\begin{array}{ccc}
i \gamma^{\mu} \partial_{\mu} - m_{n} & 0 & 0 \\
0 & i \gamma^{\mu} \partial_{\mu} -(m_{n} + m ) & 0 \\
0& 0 &i \gamma^{\mu} \partial_{\mu} -(m_{n} - m)  
\end{array}
\right)
\left(
\begin{array}{c} 
\psi_1^{(n)} \\
\tilde{\psi}_2^{(n)} \\
\tilde{\psi}_3^{(n)}
\end{array}
\right) \right. \nonumber \\ 
&& \left. 
+\frac{g}{2} (\bar{\psi}_1^{(n)}, \bar{\tilde{\psi}}_2^{(n)}, 
\bar{\tilde{\psi}}_3^{(n)})  
\left(
\begin{array}{ccc}
W^{3}_{\mu}+ \frac{\sqrt{3}B_{\mu}}{3} & W^{+}_{\mu} & W^{+}_{\mu} \\
W^{-}_{\mu} & - \frac{W^{3}_{\mu}}{2} - \frac{\sqrt{3} B_{\mu}}{6} & 
- \frac{W^{3}_{\mu}}{2} + \frac{\sqrt{3} B_{\mu}}{2}  \\
W^{-}_{\mu} & - \frac{W^{3}_{\mu}}{2} + \frac{\sqrt{3} B_{\mu}}{2} & 
- \frac{W^{3}_{\mu}}{2} - \frac{\sqrt{3} B_{\mu}}{6} 
\end{array}
\right) 
\gamma^{\mu} 
\left(
\begin{array}{c}
\psi_1^{(n)} \\
\tilde{\psi}_2^{(n)} \\
\tilde{\psi}_3^{(n)}
\end{array}
\right) \right\} \nonumber  \\ 
&&   + \ \mbox{zero-mode part}.  
\label{4Deff}
\eea
The relevant Feynman rules for our calculation can be readily 
read off from this lagrangian. 

\section{Calculation of T-parameter in 5D}
\begin{figure}[t]
 \begin{center}
  \includegraphics[width=2.5cm]{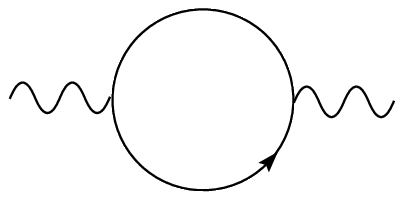}
  \hspace*{5mm}
  \includegraphics[width=2.5cm]{oblique.eps}
  \hspace*{5mm}
  \includegraphics[width=2.5cm]{oblique.eps}
  \hspace*{5mm}
  \includegraphics[width=2.5cm]{oblique.eps}
   \put(-325,-30){(1a)}
  \put(-232,-30){(1b)}
  \put(-137,-30){(1c)}
  \put(-45,-30){(1d)}
\put(-320,42){$\psi_1^{(n)}$}
\put(-320,-12){$\psi_1^{(n)}$}
\put(-228,42){$\tilde{\psi}_2^{(n)}$}
\put(-228,-12){$\tilde{\psi}_2^{(n)}$}
\put(-135,42){$\tilde{\psi}_3^{(n)}$}
\put(-135,-12){$\tilde{\psi}_3^{(n)}$}
\put(-42,42){$\tilde{\psi}_3^{(n)}$}
\put(-42,-12){$\tilde{\psi}_2^{(n)}$}
\put(-395,17){$\delta M^2_{W^3}$}
\put(-363,17){$=$}
\put(-272,17){$+$}
\put(-180,17){$+$}
\put(-85,17){$+$}
 \end{center}
\caption{One-loop diagrams contributing to 
the neutral W boson mass-squared due to the non-zero K-K modes of 
fermions. The external lines denote $W_\mu^3$ having no external momenta.} 
\label{neutralT}
\end{figure}

\begin{figure}[h]
 \begin{center}
  \includegraphics[width=2.5cm]{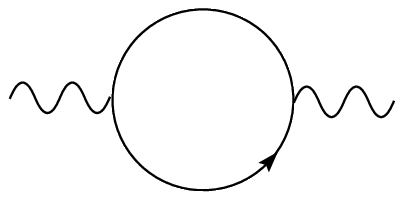}
  \hspace*{5mm}
  \includegraphics[width=2.5cm]{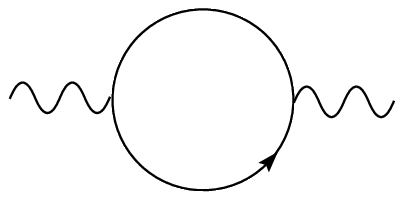}
   \put(-140,-30){(2a)}
  \put(-45,-30){(2b)}
\put(-135,42){$\psi_1^{(n)}$}
\put(-135,-12){$\tilde{\psi}_2^{(n)}$}
\put(-40,42){$\psi_1^{(n)}$}
\put(-40,-12){$\tilde{\psi}_3^{(n)}$}
\put(-215,17){$\delta M^2_{W^{\pm}}$}
\put(-178,17){$=$}
\put(-85,17){$+$}
 \end{center}
\caption{One-loop diagrams contributing to 
the charged W boson mass-squared due to the non-zero K-K modes.
The external lines denote $W_\mu^{\pm}$.} 
\label{chargedT}
\end{figure}

In this section, 
we calculate the one-loop contribution to the T-parameter 
from the matter fermions in 5D space-time. For that purpose, 
we calculate the mass-squared difference 
between neutral and charged W-bosons 
$\Delta M^2 \equiv \delta M^2_{W^3} - \delta M^2_{W^{\pm}}$. 
We first derive general formulas in the space-time $M^{D} \times S^{1}/Z_2$ 
for later use, and finally set $D = 4$. 
The contributions from non-zero K-K modes are obtained 
from the diagrams in Fig. \ref{neutralT} and Fig. \ref{chargedT}. 

Let us first calculate the diagrams contributing to 
the neutral W-boson mass-squared shown in Fig. \ref{neutralT}. 

We first note that diagrams (1a), (1b) and (1c) actually do not contribute. 
This is simply because the gauge couplings of non-zero K-K modes of the fermions are vector-like 
and therefore these diagrams are just the same as 
the quantum correction to the photon mass in ordinary QED, 
which should vanish due to the gauge invariance. 
We have confirmed this is the case by performing $D$-dimensional momentum integral 
by use of dimensional regularization for each K-K mode $n$.  
The contribution of the remaining diagram (1d) is calculated to be  
\bea
({\rm 1d}) &=& 
i\frac{g^2 N_{c}}{8}\frac{2^{D/2}}{D} \sum_{n=1}^\infty \int\frac{d^Dk}{(2\pi)^D} 
\frac{(2-D)k^2+D(m_n^2-m^2)}{[k^2-(m_{n}-m)^2][k^{2} - (m_{n}+m)^{2}]} 
\label{1d1} \nonumber \\
&=& 
i\frac{g^2 N_{c}}{8}\frac{2^{D/2}}{D} \sum_{n=1}^\infty \int_0^1 dt \int\frac{d^Dk}{(2\pi)^D} 
\frac{(2-D)k^2+D(m_n^2-m^2)}{[k^2-(m_n^2+m^2) + 2m_n m(2t-1)]^2} \nonumber \\
&=& 
-\frac{g^2 N_{c}}{4(4\pi)^{D/2}} 2^{D/2} \Gamma \left( 2-\frac{D}{2} \right) 
\sum_{n=1}^\infty \int_0^1 dt 
\frac{m_n m(2t-1)-m^2}{[m_n^2+m^2-2m_nm(2t-1)]^{2-\frac{D}{2}}},  
\label{1d2}
\eea
where $N_{c} = 3$ is the color degree of freedom. 
In the last line, we adopted the dimensional regularization 
for the D-dimensional momentum integral. 

Calculating the diagrams (2a) and (2b) contributing to the charged $W$ boson mass-squared in a similar way, 
we obtain
\bea
({\rm 2a}) + ({\rm 2b}) &=& 
i\frac{g^2 N_{c}}{4}\frac{2^{D/2}}{D} \sum_{n=1}^\infty \int\frac{d^Dk}{(2\pi)^D} 
\frac{(2-D)k^2+D(m_n^2 + m_{n}m)}{[k^2 - m_{n}^2][k^{2} - (m_{n}+m)^{2}]} 
\label{1d1} 
+ (m \to -m) \label{2a2b1} \nonumber \\
&=& 
i\frac{g^2 N_{c}}{4} \sum_{n=1}^\infty \int_0^1dt \int\frac{d^Dk}{(2\pi)^D} 
\frac{(2-D)k^2+D(m_n^2 + m_n m)}{[k^2-m_n^2 - t(2m_n m + m^2)]^2} 
\frac{2^{D/2}}{D} + (m \to -m) \nonumber \\ 
&=& 
\frac{g^2 N_{c}}{4(4\pi)^{D/2}} 2^{D/2}\Gamma\left( 2-\frac{D}{2} \right) 
\sum_{n=1}^\infty \int_0^1dt 
\frac{(2t-1)m_n m + t m^2}{[m_n^2+t(2m_nm + m^2)]^{2-\frac{D}{2}}} 
+ (m \to -m), \nonumber \\
\label{2a2b2}
\eea
where we note $(2b) = m \to -m~{\rm in}~(2a)$. 
Thus, we get the contribution of the non-zero K-K modes to the T-parameter as 
(setting $N_{c} = 3$ and using (\ref{Tdef}) 
with $\sin^{2} \theta_{W} = \frac{3}{4}$) 
\bea
T_{(n \ne 0)} &=&  
-i\frac{2\pi}{M_W^{2}}\frac{2^{D/2}}{D} 
\sum_{n=1}^\infty \int\frac{d^Dk}{(2\pi)^D} 
\frac{(2-D)k^2+D(m_n^2-m^2)}{[k^2-(m_{n}-m)^2][k^{2} - (m_{n}+m)^{2}]} 
\nonumber \\ 
&& + \left\{
i\frac{4\pi}{M_{W}^{2}} \frac{2^{D/2}}{D} 
\sum_{n=1}^\infty \int\frac{d^Dk}{(2\pi)^D} 
\frac{(2-D)k^2+D(m_n^2 + m_{n}m)}{[k^2 - m_{n}^2][k^{2} - (m_{n}+m)^{2}]} 
\label{1d1} + (m \to -m)
\right\}
\label{TKK1} \nonumber \\ 
&=& 
-i\frac{2\pi}{M_W^{2}} \frac{2^{D/2}}{D}\sum_{n=1}^\infty 
\int_0^1 dt \int \frac{d^Dk}{(2\pi)^D}
\left[
\frac{(2-D)k^2 + D(m_n^2-m^2)}{[k^2-(m_n^2+m^2) +2m_nm(2t-1)]^2} 
\right. \nonumber \\
&& \left. 
-2\left\{ \frac{(2-D)k^2+D(m_n^2 + m_n m)}{[k^2 - m_n^2 - t(2m_n m + m^2)]^2} 
+ (m \to -m) \right\}
\right] \nonumber \\
&=& 
- \frac{2^{D/2}}{(4\pi)^{D/2-1} M_W^{2}} \sum_{n=1}^\infty 
\int_0^1 dt \Gamma(2-\frac{D}{2})
\left[
\frac{m^2+m_n m(1-2t)}{[m_n^2+m^2 +2m_nm(1-2t)]^{2-D/2}} 
\right. \nonumber \\
&& \left. 
-\left\{ \frac{- m_n m + t(2m_n m + m^2)}
{[m_n^2 + t(2m_n m + m^2)]^{2-D/2}} 
+ (m \to -m) \right\}
\right]. 
\label{TKK2}
\eea
As we expect, $T_{(n \neq 0)}$ vanishes 
in the limit $m \to 0$,  
which corresponds to the limit of the custodial symmetry in our model. 

Recalling $M_W = m$ in our toy model, we have
\bea
T_{n \ne 0} 
&=& 
- \frac{2^{D/2}}{(4\pi)^{D/2-1} m^{2}} \sum_{n=1}^\infty 
\int_0^1 dt \Gamma(2-\frac{D}{2})
\left[
\frac{m^2+m_n m(1-2t)}{[m_n^2+m^2 +2m_nm(1-2t)]^{2-D/2}} 
\right. \nonumber \\
\label{Tneq0}
&& \left. 
-\left\{ \frac{- m_n m + t(2m_n m + m^2)}
{[m_n^2 + t(2m_n m + m^2)]^{2-D/2}} 
+ (m \to -m) \right\}
\right]. 
\eea


\begin{figure}[h]
 \begin{center}
  \includegraphics[width=2.5cm]{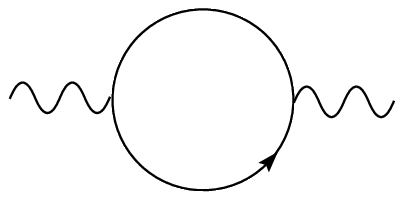}
  \hspace*{5mm}
  \includegraphics[width=2.5cm]{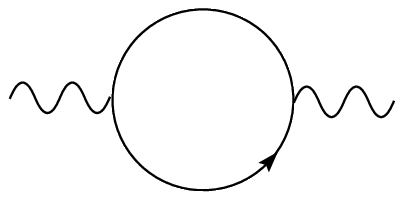}
   \put(-140,-30){(3a)}
  \put(-45,-30){(3b)}
\put(-130,42){$t$}
\put(-130,-12){$t$}
\put(-38,42){$b$}
\put(-38,-12){$b$}
\put(-215,17){$\delta M^2_{W^{3}}$}
\put(-178,17){$=$}
\put(-85,17){$+$}
 \end{center}
\caption{One-loop diagrams contributing to 
the neutral W boson mass-squared due to the zero modes of 
fermions. The external lines denote $W_\mu^3$.} 
\label{neutralT0}
\end{figure}

\begin{figure}[h]
 \begin{center}
  \includegraphics[width=2.5cm]{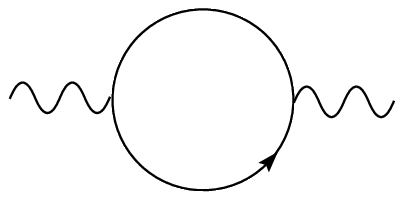}
  \put(-45,-30){(4a)}
\put(-36,42){$t$}
\put(-36,-12){$b$}
\put(-120,17){$\delta M^2_{W^{\pm}}$}
\put(-83,17){$=$}
 \end{center}
\caption{A one-loop diagram contributing to 
the charged W boson mass-squared due to the zero modes of 
fermions. The external lines denote $W_\mu^{\pm}$.} 
\label{chargedT0}
\end{figure}

In (\ref{TKK2}), the D-dimensional momentum integral was performed 
before the K-K mode sum. 
We can equally perform the K-K mode sum first. 
In this approach, 
it is convenient to include the zero-mode $(n = 0)$ contribution.  
The zero-mode contribution is calculated 
from Figs. \ref{neutralT0} and \ref{chargedT0}:  
\bea
\Delta M^2_{(n=0)} &=& 
ig^2 N_{c} \frac{2^{D/2}}{8}\frac{(2-D)}{D} 
\int \frac{d^Dk}{(2\pi)^D} \frac{m^4}{k^2(k^2-m^2)^2} \nonumber \\
&=& \frac{g^2 N_{c}}{(4\pi)^{D/2}}\frac{2^{D/2}}{8}\frac{D-4}{D}
\Gamma\left( 2 - \frac{D}{2} \right) (m^2)^{D/2-1}. 
\label{0modeT}
\eea
Let us note that the zero mode contribution just coincides 
with the half of what we obtain by setting $n = 0$, 
instead of the summation $\sum_{n > 0}$, in (\ref{1d2}) minus (\ref{2a2b2}). 
Thus, by using (\ref{1d2}) minus (\ref{2a2b2}) 
the whole contribution to the T-parameter can be neatly written 
in terms of $\sum_{n=-\infty}^{\infty}$ as 
\bea
\Delta M^2 &=& i\frac{g^2 N_{c}}{16}\frac{2^{D/2}}{D}\sum_{n=-\infty}^\infty 
\int \frac{d^Dk}{(2\pi)^D} \nonumber \\
&\times& \left[
\frac{(2-D)k^2 + D(m_n^2 - m^2)}{[k^2 - (m_n - m)^2][k^2 - (m_n + m)^2]}
-4 \frac{(2-D)k^2 + D(m_n^2 + m_{n}m)}{[k^2 - m_n^2][k^2 - (m_n + m)^2]}
\right]. 
\label{Tall}
\eea
We rewrite (\ref{Tall}) as follows, 
\bea
\Delta M^2 &=& i\frac{g^2 N_{c}}{16}\frac{2^{D/2}}{D}\int_0^1dt \int 
\frac{d^Dk}{(2\pi)^D}\sum_{n=-\infty}^\infty
\left[
\frac{2D}{k^2-m_n^2} + \frac{D}{k^2-(m_n+m)^2} \right. \nonumber \\
&+& \left. \frac{2(k^2-Dm^2)}{[k^2-(m_n + (2t-1)m)^2 +4t(t-1)m^2]^2} 
-\frac{2(4k^2 - Dm^2)}{[k^2-(m_n + tm)^2 +t(t-1)m^2]^2}
\right]. \nonumber \\
\eea
Using the formulas, 
\bea
\sum_{n=-\infty}^\infty \frac{1}{x^2 + (a + 2n \pi)^2} 
&=& \frac{\sinh x}{2x(\cosh x - \cos a)}, 
\label{formula1} \\
\sum_{n=-\infty}^\infty \frac{1}{[x^2 + (a + 2n \pi)^2]^2} 
&=& 
-\frac{1}{2x}\frac{\partial}{\partial x}\sum_{n=-\infty}^\infty 
\frac{1}{x^2 + (a + 2n \pi)^2} \nonumber \\
&=& -\frac{1}{4x} \frac{\partial}{\partial x} \left[ \frac{\sinh x}{x(\cosh x - \cos a)} \right], 
\label{formula2}
\eea
we obtain the expression for $\Delta M^2$ after taking the sum over $n$ 
\bea
\Delta M^2 &=& - \frac{g^2 N_{c}}{16} \frac{2^{D/2}}{D}L^{2-D} 
\int_0^1 dt \int 
\frac{d^D \rho}{(2\pi)^D} 
\left[
-\frac{D \sinh \rho}{\rho(\cosh \rho -1)} 
-\frac{D \sinh \rho}{2\rho(\cosh \rho -\cos \alpha)} \right. \nonumber \\
&& \left. 
+2(\rho^2 + D \alpha^2) 
\left( \frac{1}{4\rho} \right) 
\frac{\partial}{\partial \rho} 
\left\{
\frac{1}{\sqrt{\rho^2 + 4t(1-t)\alpha^2}}
\right. \right. \nonumber \\
&& \left. \left. 
+\frac{1}{\sqrt{\rho^2 + 4t(1-t)\alpha^2}}
\left(
\frac{\sinh \sqrt{\rho^2 + 4t(1-t)\alpha^2}}
{\cosh \sqrt{\rho^2 + 4t(1-t)\alpha^2}-\cos[(2t-1)\alpha]} 
 -1 \right)
\right\} \right. \nonumber \\
&& \left. -2(4\rho^2 + D \alpha^2) 
\left( \frac{1}{4\rho} \right) 
\frac{\partial}{\partial \rho} 
\left\{
\frac{1}{\sqrt{\rho^2 + t(1-t)\alpha^2}}
\right. \right. \nonumber \\
&& \left. \left. 
+\frac{1}{\sqrt{\rho^2 + t(1-t)\alpha^2}}
\left(
\frac{\sinh \sqrt{\rho^2 + t(1-t)\alpha^2}}
{\cosh \sqrt{\rho^2 + t(1-t)\alpha^2}-\cos (t \alpha)} -1 
\right)
\right\}
\right] 
\label{summedT}
\eea
where $L \equiv 2\pi R, \rho^{\mu} \equiv L k_{E}^{\mu}$, 
with $k_{E}^{\mu}$ being the Euclidean momentum, and 
$\alpha \equiv Lm$ is the ``Aharanov-Bohm" phase.

Since the quantities in the integrand,  
\bea
\frac{1}{\sqrt{\rho^2 + 4t(1-t)\alpha^2}}
\left(
\frac{\sinh \sqrt{\rho^2 + 4t(1-t)\alpha^2}}
{\cosh \sqrt{\rho^2 + 4t(1-t)\alpha^2}-\cos[(2t-1)\alpha]} 
 -1 \right) 
\eea
etc., do not have UV nor IR divergence when it is multiplied by 
$\rho^{D-1}\frac{(\rho^2 + D \alpha^2)}{\rho}$, etc.,   
it is useful to perform the integration by parts to obtain 
\bea
T &=& T_{({\rm div})} + T_{({\rm sc})}, \\
T_{({\rm div})} &=& \frac{\pi}{\alpha^{2}}\frac{2^{D/2}}{D} 
L^{4-D} \int_0^1 dt \int\frac{d^D\rho}{(2\pi)^D} \nonumber \\
&& \times \left[
-\frac{3D}{2\rho} 
- \frac{\rho^2 + D \alpha^2}{2[\rho^2 + 4t(1-t)\alpha^2]^{3/2}} 
+\frac{4\rho^2 + D \alpha^2}{2[\rho^2 + t(1-t)\alpha^2]^{3/2}}
\right] \nonumber \\ 
&=& -\pi \frac{2^{\frac{3}{2}D-3}}{(4\pi)^{D/2}}
\frac{(1-2^{3-D})(D-1)}{D(3-D)}
\frac{\Gamma(\frac{5-D}{2})\Gamma(\frac{D-1}{2})^2}
{\Gamma(\frac{3}{2})\Gamma(D-1)}L^{4-D}\alpha^{D-3}, 
\label{divT}\\
T_{({\rm sc})} &=& 
\frac{\pi}{\alpha^{2}}\frac{2^{\frac{D}{2}}}{D}
L^{4-D} 
\int_0^1dt \int \ \frac{d^{D}\rho}{(2\pi)^{D}} \nonumber \\
&& \times \left[
-\frac{D}{\rho} \left( \frac{\sinh \rho}{\cosh \rho -1} - 1 \right) 
-\frac{D}{2\rho}\left( \frac{\sinh \rho}{\cosh \rho -\cos \alpha} - 1 \right) 
\right. \nonumber \\
&& \left. 
-\frac{D}{2} 
\frac{\left(1 + (D-2) \frac{\alpha^2}{\rho^2} \right)}
{\sqrt{\rho^2 + 4t(1-t)\alpha^2}}
\left(
\frac{\sinh \sqrt{\rho^2 + 4t(1-t)\alpha^2}}
{\cosh \sqrt{\rho^2 + 4t(1-t)\alpha^2}-\cos[(2t-1)\alpha]}  -1 \right)
\right. \nonumber \\
&& \left. 
+\frac{D}{2}
\frac{\left(4 + (D-2) \frac{\alpha^2}{\rho^2} \right)}
{\sqrt{\rho^2 + t(1-t)\alpha^2}}
\left(
\frac{\sinh \sqrt{\rho^2 + t(1-t)\alpha^2}}
{\cosh \sqrt{\rho^2 + t(1-t)\alpha^2}-\cos (t \alpha)} -1 
\right) 
\right] 
\label{scTexact}, 
\eea
where $T_{({\rm div})}$ denotes a possibly divergent part, 
a part which seems to be UV divergent relying on a naive power counting, 
though it is actually finite in 5D space-time (D=4). 
$T_{({\rm sc})}$ denotes an apparently super-convergent part. 
(\ref{scTexact}) is the exact formula, 
valid for arbitrary $m$ ($\alpha$), and can be 
evaluated by performing the convergent integrals, 
if necessary by numerical computation.    
 
Now let us discuss the T-parameter in 5D space-time 
by taking the limit $D \to 4$. 
We first utilize the approach to carry out the momentum integral 
before taking the mode sum. 
In the limit $D \to 4$, 
the contribution of $n \neq 0$ modes (\ref{Tneq0}) reduces to  
\bea
T_{(n \ne 0)} (\mbox{5D}) &=& - \frac{1}{\pi m^{2}} 
\sum_{n=1}^\infty \int_0^1 dt \times \nonumber \\
&&\left[
(-(1-2t)m_nm + t m^2) \ln[m_n^2+ t(2m_nm + m^2)] 
\right. \nonumber \\
&&\left. + ((1-2t)m_nm + t m^2)\ln[m_n^2 + t(-2m_nm + m^2)] 
\right. \nonumber \\
&& \left. - (m^2 + (1-2t)m_n m)\ln[m_n^2+m^2+2(1-2t)m_nm]
\right], 
\eea  
where the pole term of $\Gamma (2 - \frac{D}{2})$ is known to vanish, as 
$\int_{0}^{1} \ (1-2t) dt = 0$. 
Therefore, the T-parameter in 5D turns out to be finite. 

The finite part can be explicitly evaluated 
if we adopt a reasonable approximation, 
$m \ll \frac{1}{R}$, i.e. $m/m_{n} \ll 1$.  
Thus, expanding the integrand in the powers of $m/m_n$ 
up to ${\cal O}((m/m_n)^4)$ 
and integrating over $t$, we obtain
\bea
T_{(n \ne 0)} (\mbox{5D}) \simeq  
\frac{2}{5\pi m^{2}} \sum_{n=1}^\infty \frac{m^4}{m_n^2} 
= \frac{\pi}{15}(mR)^2, 
\label{5DT}
\eea
where $\sum_{n=1}^\infty n^{-2} = \zeta(2) = \pi^2/6$ is used. 
The fact that the leading order term of each K-K mode's contribution is proportional to $\frac{m^2}{m_n^2}$, 
corresponding to the leading contribution of ${\cal O}(\frac{m^{4}}{m_{n}^{2}})$ in $\Delta M^{2}$, 
is the consequence of that the dominant contribution of 
the heavy $n \neq 0$ K-K modes to the T parameter ($\Delta M^{2}$)
is obtained by the insertion of VEV for the Higgs field $\phi$ 
in the 4D operator with mass dimension six, responsible for the parameter, 
$(\phi^\dag D_\mu \phi)(\phi^\dag D^\mu \phi)$, 
accompanied by the coefficient suppressed by $1/m_{n}^{2}$ 
(the ``decoupling" of $n \neq 0$ K-K modes). 
The effects of the operators with higher mass dimensions are further suppressed.  

This finite value of T-parameter can be also derived from the second approach 
where K-K mode sum is taken before the momentum integration, discussed above, 
which is useful to see the structure of UV divergence. 
Namely, for the 5D case (D=4), 
we find that the possibly divergent part (\ref{divT}) becomes 
\bea
T_{({\rm div})} (\mbox{5D}) =  
\frac{1}{4\pi m^{2}} \frac{3\pi^2}{8}(mR)m^2 
\label{5DdivT}
\eea
and is actually finite. It is found to be proportional to $mR$. 
In order to obtain the result consistent with (\ref{5DT}), 
this term should be canceled by a term in the superconvergent part.  
We can see that this is indeed the case. 
After some lengthy calculations\footnote{The details of calculation 
is explained in Appendix A.}, we get the superconvergent part 
for the 5D case ($D=4$): 
\bea
T_{({\rm sc})}(\mbox{5D}) \simeq 
\frac{1}{4\pi m^{2}} 
\left[
m^2-\frac{3\pi^2}{8}(mR)m^2 + \frac{4\pi^2}{15}(mR)^2 m^2
\right]. 
\label{5DscT}
\eea
Combining (\ref{5DdivT}) and (\ref{5DscT}), 
we obtain 
\bea
T(\mbox{5D}) \simeq \frac{1}{4\pi}
\left(
1 + \frac{4\pi^2}{15}(mR)^2
\right). 
\label{5DTfromsum}
\eea
One can see that the $mR$ term (\ref{5DdivT}) from the possibly divergent part 
is exactly canceled by the $mR$ term from the superconvergent part 
(\ref{5DscT}). 
The constant term in the bracket in (\ref{5DTfromsum}) is known to coincide with the zero mode contribution 
(\ref{0modeT}) with $D=4$. 
The remaining $(mR)^2$ term agrees with the finite result 
of non-zero K-K mode contribution (\ref{5DT}), 
which was calculated by performing the momentum integral before taking the mode sum.

\section{Calculation of S-parameter in 5D}
\begin{figure}[h]
 \begin{center}
  \includegraphics[width=2cm]{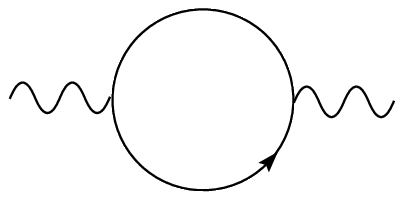}
  \hspace*{5mm}
  \includegraphics[width=2cm]{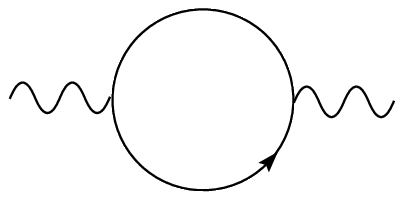}
  \hspace*{5mm}
  \includegraphics[width=2cm]{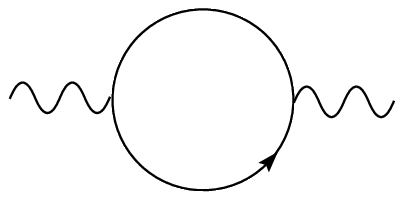}
  \hspace*{5mm}
  \includegraphics[width=2cm]{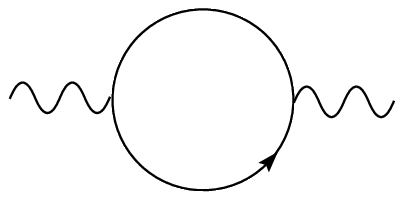}
  \hspace*{5mm}
  \includegraphics[width=2cm]{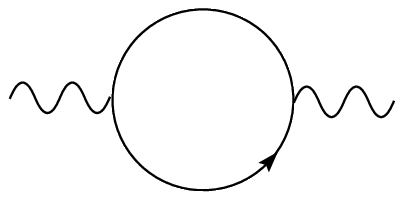}
   \put(-354,-30){(S1)}
  \put(-274,-30){(S2)}
  \put(-194,-30){(S3)}
  \put(-114,-30){(S4)}
  \put(-37,-30){(S5)}
\put(-348,35){$\psi_1^{(n)}$}
\put(-348,-12){$\psi_1^{(n)}$}
\put(-270,35){$\tilde{\psi}_2^{(n)}$}
\put(-270,-12){$\tilde{\psi}_2^{(n)}$}
\put(-190,35){$\tilde{\psi}_3^{(n)}$}
\put(-190,-12){$\tilde{\psi}_3^{(n)}$}
\put(-110,35){$\tilde{\psi}_2^{(n)}$}
\put(-110,-12){$\tilde{\psi}_3^{(n)}$}
\put(-34,35){$\tilde{\psi}_3^{(n)}$}
\put(-35,-12){$\tilde{\psi}_2^{(n)}$}
\put(-405,60){$i\Pi'_{{\rm 3Y}}p^2g_{\mu\nu} + \cdots$}
\put(-400,15){$=$}
\put(-310,13){$+$}
\put(-230,13){$+$}
\put(-150,13){$+$}
\put(-70,13){$+$}
 \end{center}
\caption{One-loop diagrams contributing to 
the self-energy between $W^3_{\mu}$ and $B_{\mu}$ from the non-zero K-K modes of fermions.} 
\label{KKS}
\end{figure}

\begin{figure}[h]
 \begin{center}
  \includegraphics[width=2.5cm]{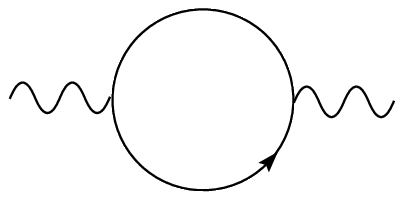}
  \hspace*{5mm}
  \includegraphics[width=2.5cm]{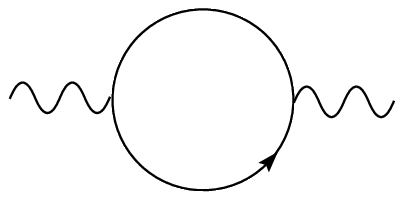}
   \put(-140,-30){(S6)}
  \put(-45,-30){(S7)}
\put(-130,42){$t$}
\put(-130,-12){$t$}
\put(-38,42){$b$}
\put(-38,-12){$b$}
\put(-275,17){$i\Pi'_{{\rm 3Y}}p^2g_{\mu\nu} + \cdots$}
\put(-178,17){$=$}
\put(-85,17){$+$}
 \end{center}
\caption{One-loop diagrams contributing to 
the self-energy between $W^3_{\mu}$ and $B_{\mu}$ from the zero mode of fermions.} 
\label{S0mode}
\end{figure}

In this section, we calculate one-loop contribution to the S-parameter, 
which is calculated from the coefficient 
$\Pi'_{{\rm 3Y}}$ of $p^{2} g_{\mu \nu}$ 
term in the self-energy between two neutral gauge bosons 
$W_\mu^3$ and $B_{\mu}$, 
$\Pi_{{\rm 3Y}}(p^2)_{\mu\nu} = \Pi'_{{\rm 3Y}}p^2g_{\mu\nu} + \cdots$. 
The diagrams we have to calculate are listed in Fig. \ref{KKS} 
and Fig. \ref{S0mode} where the former shows non-zero K-K mode contribution  
and the latter does zero mode contribution, respectively. 
Let us first calculate one-loop contribution to the S-parameter 
due to the $n \neq 0$ K-K modes, 
by performing momentum integration first by use of dimensional regularization. 
The Taylor expansion of the first diagram (S1) 
in terms of external momentum $p_{\mu}$ yields 
a contribution at the order ${\cal O} (p^{2})$,  
\bea
\Pi_{{\rm 3Y}}^{(S1)}(p^2)_{\mu\nu} 
&\simeq& i\frac{\sqrt{3}g^2 N_{c}}{72}2^{D/2}\int\frac{d^Dk}{(2\pi)^D} 
\nonumber \\
&& \times \sum_{n=1}^\infty 
\left[
\left(
\frac{3}{(k^2 - m_n^2)^2} 
- \frac{4k^2}{D(k^2 - m_n^2)^3}
\right)p^2g_{\mu\nu} 
-\frac{2p_\mu p_\nu}{(k^2 - m_n^2)^2}
\right] 
\label{s1p2} \\
&=& -\frac{\sqrt{3}g^2 N_{c}}{36}\frac{2^{D/2}}{(4\pi)^{D/2}}
\Gamma\left(2 - \frac{D}{2} \right) \sum_{n=1}^\infty 
(m_{n}^2)^{\frac{D}{2}-2}(p^2g_{\mu\nu}-p_\mu p_\nu) 
\label{s1dr} \\ 
&=& 
i\frac{\sqrt{3}g^2 N_{c}}{36} 2^{D/2}
\int \frac{d^Dk}{(2\pi)^D}\sum_{n=1}^\infty 
\frac{1}{(k^2 - m_n^2)^2}(p^2g_{\mu\nu}-p_\mu p_\nu)
\label{s1}
\eea
where $\simeq$ means that 
the only ${\cal O}(p^2)$ terms relevant for the S-parameter are picked up. 
Superficially, (\ref{s1p2}) and (\ref{s1}) look different, 
but they can be identified through dimensional regularization (\ref{s1dr}). 
Obtained result satisfies CVC relation $p^\mu \Pi_{{\rm 3Y}\mu\nu}^{(S1)}=0$, 
which is again the reflection of the fact that the gauge couplings 
of $n \neq 0$ fermions are vector-like, 
just as in the ordinary QED. The CVC relation also holds for 
the diagrams (S2) and (S3), as we will see below.

The remaining diagrams due to $n \neq 0$ modes can be calculated 
in a similar manner:  
\bea
\Pi_{{\rm 3Y}}^{(S2)+(S3)}(p^2)_{\mu\nu} 
&\simeq& 
-\frac{\sqrt{3}g^2 N_{c}}{144}\frac{2^{D/2}}{(4\pi)^{D/2}} \nonumber \\
&& \times \Gamma\left( 2-\frac{D}{2} \right)
\sum_{n=1}^\infty [(m_n + m)^2]^{\frac{D}{2}-2}
(p^2g_{\mu\nu} - p_\mu p_\nu) + (m \to -m) \nonumber \\
&=& 
i\frac{\sqrt{3}g^2 N_{c}}{144} 2^{D/2}
\int\frac{d^Dk}{(2\pi)^D}
\sum_{n=1}^\infty \frac{1}{[k^2 - (m_n + m)^2]^2}
(p^2g_{\mu\nu} - p_\mu p_\nu) \nonumber \\
&& +(m \to -m), 
\label{s2s3}
\eea
\bea
\Pi_{{\rm 3Y}}^{(S4)+(S5)}(p^2)_{\mu\nu} 
&\simeq& 
\frac{\sqrt{3}g^2 N_{c}}{8}\frac{2^{D/2}}{(4\pi)^{D/2}} 
\int_0^1 dt t(1-t) \nonumber \\
&& \times \sum_{n=1}^\infty 
\left[
\Gamma \left(2-\frac{D}{2} \right)
\frac{1}{[m_n^2+m^2+2(2t-1)m_n m]^{2-\frac{D}{2}}}
(p^2g_{\mu\nu} - p_\mu p_\nu) \right. \nonumber \\
&& \left. -\Gamma \left(3-\frac{D}{2} \right)
\frac{(2t-1)[m_n + (2t-1)m]m + 4t(1-t)m^2}
{[m_n^2 + m^2 + 2(2t-1)m_n m]^{3-\frac{D}{2}}}p^2g_{\mu\nu}
\right] \nonumber \\
&& +(m \to -m) \nonumber \\
&=&
-i\frac{\sqrt{3}g^2 N_{c}}{8} 2^{D/2} 
\int_0^1 dt t(1-t) \int \frac{d^Dk}{(2\pi)^D} \nonumber \\
&& \times \sum_{n=1}^\infty 
\left[
\frac{1}{[k^2 - (m_n +(2t-1)m)^2-4t(1-t)m^2]^2}
(p^2g_{\mu\nu} - p_\mu p_\nu) \right. \nonumber \\
&& \left. +2
\frac{(2t-1)[m_n + (2t-1)m]m + 4t(1-t)m^2}
{[k^2 - (m_n + (2t-1) m)^2-4t(1-t)m^2]^3} p^2g_{\mu\nu}
\right] \nonumber \\
&&+(m \to -m)
\label{s4s5}
\eea
where we made use of the property that 
the diagrams (S3) and (S5) are the same diagrams as (S2) and (S4), 
respectively, if we replace $m$ by $-m$.

Thus, non-zero K-K mode contributions to the S-parameter (\ref{Sdef}), 
after the momentum integral, are summarized as follows (with $\tan \theta_{W} = \sqrt{3}$ and $N_{c} = 3$),  
\bea
&&S_{(n \neq 0)} \nonumber \\
&=& \frac{\pi}{3}2^{D/2}\frac{\Gamma(2-\frac{D}{2})}{(4\pi)^{D/2}} 
\sum_{n=1}^\infty \left[
\frac{4}{(m_n^2)^{2-D/2}} +\frac{1}{[(m_n + m)^2]^{2-D/2}} 
+\frac{1}{[(m_n - m)^2]^{2-D/2}} \right. \nonumber \\
&& \left. -18 \int_0^1 dt t(1-t) \right. \nonumber \\
&& \left. \times \left\{
\frac{1}{[m_n^2 + m^2 + 2(2t-1)m_n m]^{2-D/2}}
+\frac{1}{[m_n^2 + m^2 - 2(2t-1)m_n m]^{2-D/2}}
\right\}
\right] \nonumber \\
&&+6\pi 2^{D/2}\frac{\Gamma(3-\frac{D}{2})}{(4\pi)^{D/2}} 
\sum_{n=1}^\infty \int_0^1 dt 
\left[
\frac{t(1-t)[(2t-1)(m_n+(2t-1)m)m+4t(1-t)m^2]}
{[m_n^2 + m^2 + 2(2t-1)m_n m]^{3-D/2}} \right. \nonumber \\
&& \left. +\frac{
t(1-t)[(2t-1)(-m_n+(2t-1)m)m+4t(1-t)m^2]
}{[m_n^2 + m^2 - 2(2t-1)m_n m]^{3-D/2}}
\right]. 
\label{Sdr}
\eea
We can easily check that UV divergence (for the case of $D=4$) 
is cancelled out for a fixed K-K mode as 
\bea
\left[ 2 + 1 - 18 \int_0^1 dt t(1-t) \right] 
\times ({\rm log~divergence}) = 0. 
\eea

Next we take another approach, 
i.e. we perform the K-K mode sum before the momentum integral. 
First let us consider zero mode contributions. They are given by
\bea
\Pi^{(S6)}_{{\rm 3Y}}(p^2)_{\mu\nu} 
&=& i\frac{\sqrt{3}g^2 N_{c}}{24}2^{D/2}\int_0^1 dt 
\int\frac{d^Dk}{(2\pi)^D} \frac{(\frac{2-D}{D}k^2 + t(1-t)p^2)g_{\mu\nu}
- 2t(1-t) p_\mu p_\nu}{[k^2 + t(1-t)p^2]^2}, \nonumber \\ 
\\
\Pi^{(S7)}_{{\rm 3Y}}(p^2)_{\mu\nu} 
&=& -i\frac{\sqrt{3}g^2 N_{c}}{24}2^{D/2} \times \nonumber \\
&& \int_0^1 dt 
\int\frac{d^Dk}{(2\pi)^D} \frac{(\frac{2-D}{D}k^2 + t(1-t)p^2 -2m^2)g_{\mu\nu}
- 2t(1-t) p_\mu p_\nu}{[k^2 + t(1-t)p^2 - m^2]^2}. 
\label{0modeS}
\eea
Noticing the fact 
\bea
\Pi^{(S6)}_{{3Y}}(p^2)_{\mu\nu} &=& \frac{1}{2} 
\Pi^{(S1)}_{{\rm 3Y}}(p^2)_{\mu\nu}~{\rm with}~m_n = 0, \\
\Pi^{(S7)}_{{3Y}}(p^2)_{\mu\nu} &=& 
\Pi^{(S2)}_{{\rm 3Y}}(p^2)_{\mu\nu} 
+ \Pi^{(S4)}_{{\rm 3Y}}(p^2)_{\mu\nu}~{\rm with}~m_n = 0, 
\eea
the sum of all K-K mode contributions can be written as
\bea
\Pi'_{{\rm 3Y}}(0) &=& 
i\frac{\sqrt{3}g^2 N_{c}}{144}2^{D/2}\int \frac{d^Dk}{(2\pi)^D} 
\sum_{n=-\infty}^\infty 
\left[
\frac{2}{(k^2-m_n^2)^2} + \frac{1}{[k^2-(m_n + m)^2]^2} 
\right. \nonumber \\
&& \left. -18\int_0^1 dt t(1-t) 
\left\{
\frac{1}{[k^2 - (m_n + (2t-1)m)^2 - 4t(1-t)m^2]^2} 
\right. \right. \nonumber \\
&& \left. \left. + 2 \frac{(2t-1)[m_n + (2t-1)m]m + 4t(1-t)m^2}
{[k^2 - (m_n + (2t-1)m)^2 -4t(1-t)m^2]^3}
\right\}
\right]. 
\label{allS}
\eea
The first term in the right hand side of (\ref{allS}) contains 
IR divergence for $n=0$ (and for $D=4$), which reflects the IR divergence 
we have when we take the limit $m_t \to 0$ in the $\ln(m_t/m_b)$ term of 
the ordinary $(t, b)$ doublet contribution to the S-parameter. The 
IR divergence will be cured below. 
  
In addition to (\ref{formula1}) and (\ref{formula2}), 
using the formulas 
\bea
\sum_{n=-\infty}^\infty \frac{1}{[x^2 + (a + 2n \pi)^2]^3} 
&=& 
\frac{1}{8}\frac{1}{x} 
\frac{\partial}{\partial x} \frac{1}{x} 
\frac{\partial}{\partial x} \sum_{n=-\infty}^\infty 
\frac{1}{x^2 + (a + 2n \pi)^2} \nonumber \\  
&=&
\frac{1}{16}\frac{1}{x} 
\frac{\partial}{\partial x} \frac{1}{x} 
\frac{\partial}{\partial x}  
\left[ 
\frac{\sinh x}{x(\cosh x - \cos a)}
\right], 
\eea
etc., we can make the K-K mode sum explicitly to obtain 
\bea
S &=& S_{({\rm div})} 
+ S_{({\rm sc})} \\ 
S_{({\rm div})} &=& \frac{\pi}{3}2^{D/2}L^{4-D} 
\int\frac{d^D\rho}{(2\pi)^D} 
\left\{ \frac{3}{4}\frac{1}{\rho^{3}} -18 \int_0^1 dt t(1-t) \times 
\right. \nonumber \\ 
&& \left. 
\left( \frac{1}{4 (\rho^{2} + 4t(1-t) \alpha^{2})^{3/2}} 
- \frac{3t(1-t)\alpha^{2}}{2} 
\frac{1}{(\rho^{2} + 4t(1-t) \alpha^{2})^{5/2}} \right) \right\} \nonumber \\ 
&=& \frac{9\pi 2^{3D/2-5}}{(4\pi)^{D/2}\Gamma(5/2)}
\frac{D-1}{D-3}\frac{\Gamma \left(\frac{5-D}{2} \right) 
\Gamma \left(\frac{D+1}{2} \right)^2}{\Gamma(D+1)}(2\pi R)m^{D-3}. 
\label{divS} \\  
S_{({\rm sc})}  &=& \frac{\pi}{3} 
2^{D/2}(D-2) L^{4-D} \int \frac{d^D \rho}{(2\pi)^D} \times \nonumber \\
&&\left[
\frac{1}{2\rho^3}\left(\frac{\sinh \rho}{\cosh \rho -\cos \epsilon}-1 \right) 
+\frac{1}{4\rho^3} 
\left(
\frac{\sinh \rho}{\cosh \rho - \cos \alpha} - 1 
\right) \right. \nonumber \\
&& \left. +\frac{9}{8}\int_0^1 dt 
\left\{
\left(
2t(1-t)-1
\right)\frac{1}{\rho^2}
+2t(1-t)\alpha^2 \frac{D-4}{\rho^4}
\right\} \times \right. \nonumber \\
&& \left. \frac{1}{\sqrt{\rho^2+4t(1-t)\alpha^2}} 
\left(
\frac{\sinh \sqrt{\rho^2 + 4t(1-t) \alpha^2}}
{\cosh \sqrt{\rho^2 + 4t(1-t) \alpha^2} - \cos[(2t-1)\alpha]}-1
\right)
\right], \nonumber 
\label{summedS}\\
\eea
where $\epsilon$ was introduced in the ``super-convergent" part 
$S_{({\rm sc})}$ to avoid the IR divergence due to $m_{t} = 0$ in our model. 
(The zero-mode contribution due to $(t, b)$ doublet is well-known 
and is not of our main interest in this work, anyway.)

Now, we discuss the one-loop contribution to the S-parameter in 5D space-time.
Here we adopt the result of the approach to perform the momentum integral first.  
We have already seen that the coefficient of pole term in (\ref{Sdr}) disappears. 
Therefore, one-loop contribution to the S-parameter is also finite in 5D case.  
Then, the remaining finite part in (\ref{Sdr}) can be obtained 
by expanding the logarithmic terms up to ${\cal O}((m/m_n)^2)$, 
as was done in the calculation of the T-parameter,   
and also evaluating the finite term proportional to $\Gamma(3-\frac{D}{2})$: 
\bea
S(5D) \simeq \frac{\pi}{3} \frac{1}{(2\pi)^2}
\sum_{n=1}^\infty 
\left(
\frac{28}{5} +\frac{18}{5}
\right)
\left(\frac{m}{m_n} \right)^2 
= \frac{23\pi}{180}(mR)^2
\label{finiteS}
\eea
where $28/5$ part comes from the logarithmic terms, and 
$18/5$ part is due to the $\Gamma(3-\frac{D}{2})$ term. 
$\sum_{n=1}^\infty 1/n^2 = \pi^2/6$ is used in the last equality. 
The behavior of $(m/m_{n})^2$ of each $n \neq 0$ K-K mode's contribution is consistent with what we expect from the 4D gauge invariant operator with mass dimension 6 responsible for the S-parameter,  
$(\phi^\dag W_{\mu \nu}^a \frac{\tau^a}{2}\phi) B^{\mu \nu}$, whose coefficient is suppressed by  
$m_n^{-2}$ as the result of the decoupling of massive $n \neq 0$ K-K modes.  

\section{The S and T parameters in higher than 5 dimensional space-time} 
In the previous sections, we have shown that 
one-loop contributions to the S and T parameters are finite 
in the gauge-Higgs unification scenario in 5D space-time. 
Here, we would like to clarify whether these parameters are finite or not 
in the cases higher than 5 dimensions. 

Before discussing this issue, let us recall 
why Higgs mass in the gauge-Higgs unification is finite. 
In the gauge-Higgs unification, the Higgs field is identified with the zero mode 
of extra component of gauge field in higher dimensional gauge theories. 
This implies that the local mass term for Higgs 
$\frac{1}{2} m^2 A_5^2$ (for 5D case) is strictly forbidden 
by the higher dimensional local gauge invariance. 
Although the Higgs mass is induced 
by the effect of Wilson-loop (A-B) phase, 
it is a non-local (global) operator.    
Therefore, Higgs mass in the gauge-Higgs unification is free from UV divergence. 

Then a question we should ask is whether there are 
local gauge invariant operators 
with respect to the higher dimensional gauge field $A_{M}$, 
which are responsible for the S and T parameters. 
Let us recall that in 4D space-time 
these parameters are given by the coefficients of 
dimension six operators such as 
$(\phi^\dag W_{\mu \nu} \phi)B^{\mu \nu}$ for S-parameter and 
$(\phi^\dag D_\mu \phi)(\phi^\dag D^\mu \phi)$ for T-parameter. 
Thus the operator should contain these dimension six operators 
when reduced to the 4D theory. 
At the first glance, such operators do not seem to exist, 
since the operators obtained by replacing the Higgs doublet $\phi$ 
by $A_i$ ($i$: the index to denote extra space component) 
contradict with the shift symmetry under the higher dimensional gauge transformation $A_i \to A_i + {\rm const} 
\ ({\rm for~Abelian~theories})$,  
just as in the case of Higgs mass-squared. 
Therefore, we may tend to conclude that S and T parameters 
in gauge-Higgs unification become finite. 
However, this argument is too naive and not correct: 
we find an operator to describe these parameters.  

To see this, we first note that the contribution of heavy K-K states 
should be dominated 
by the gauge invariant operators with the lowest mass dimension. 
The contributions of the operators with higher mass dimension 
will be suppressed further by the inverse powers of 
the compactification scale $M_{c} \equiv 1/R$; 
the ``decoupling" of the heavy K-K modes. 
(As for the ``non-decoupling" contributions of the zero modes $(t, b)$, 
such operators will equally contribute.) 
Thus we focus on the gauge invariant operators 
with respect to $A_{M}$ with mass dimension 6 
(when $A_{M}$ is replaced by 4D field with mass dimension one). 

Interestingly enough, such operator is unique: 
\bea   
{\rm Tr}[(D_L F_{MN})(D^L F^{MN})]. 
\eea
Let us note that by use of the Bianchi identity, 
other possible operators all reduce to this one. In fact, 
\bea 
&&{\rm Tr}[(D_L F_{MN})(a  D^L F^{MN} + b D^M F^{NL} + c D^N F^{LM} )] \nonumber \\ 
&& = \left(a - \frac{b+c}{2} \right) {\rm Tr}[(D_L F_{MN})(D^L F^{MN})], 
\eea 
for arbitrary constants $a, b$ and $c$. 

As far as there exist operators to describe the parameters, 
there is no reason for the S and T parameters to be UV finite 
in higher dimensional space-time. 
On the other hand, the fact that the S and T parameters are 
both described by a coefficient of a single operator means 
that the UV divergences appearing in the parameters 
are no longer independent of each other, but should be mutually related. 
In other words, 
if we take a specific linear combination of the S and T parameters, 
the 1-loop contribution to the combination should be finite, 
although these parameters themselves are divergent. 
It is important to note that the operator is uniquely determined 
just by the higher dimensional gauge symmetry. 
Thus the ratio of the coefficients in the linear combination should be 
independent of the detail of the matter content of the theory. 

To find out the specific linear combination, 
let us explicitly write down the relevant 
operator in terms of 4D gauge fields and the VEV of Higgs doublet,  
\bea
{\rm Tr}[(D_L F_{MN})(D^L F^{MN})] 
&\supset& 
4(g\langle A_5 \rangle)^4 
\left[
(W_\mu^3)^2 + \frac{1}{4}\left\{ (W_\mu^1)^2 + (W_\mu^2)^2 \right\}
\right] \nonumber \\
&&+\sqrt{3}(g \langle A_5 \rangle)^2 
(\partial_\mu W_\nu^3 - \partial_\nu W_\mu^3)
(\partial^\mu B^\nu - \partial^\nu B^\mu) \nonumber \\
&&+2\sqrt{3}(g\langle A_5 \rangle)^2 
(\partial_\mu W_\nu^3)(\partial^\mu B^\nu) \\
&=& \frac{1}{2}(8m^4)(W_\mu^3)^2 +(2m^4)W_\mu^+ W^{-\mu} \nonumber \\
&&+2\sqrt{3}m^2(p^2g_{\mu\nu} - p_\mu p_\nu)W^{3\mu}B^\nu 
+2\sqrt{3} m^2p^2g_{\mu\nu}W^{3\mu}B^\nu \nonumber \\
\eea
where the partial integration is carried out in the last equality and 
the transformation into the momentum space and $m = g \langle A_5 \rangle$ 
are understood. We can readily read off the contributions of 
an operator $C{\rm Tr}[(D_LF_{MN})(D^LF^{MN})] 
\ (C: \mbox{constant})$ to $\Delta M^{2}$ and $\Pi'_{3Y}$ as 
\bea
C{\rm Tr}[(D_LF_{MN})(D^LF^{MN})] \to 
\left\{
\begin{array}{l}
\Delta M^2 = 6C m^4 \\
\Pi'_{3Y} = 4\sqrt{3}C m^2, 
\end{array}
\right. 
\label{ratio}
\eea
Thus, we can expect that the linear combination 
$\Pi'_{3Y} -\frac{2}{\sqrt{3}m^2}\Delta M^2$ is free from UV divergence, since it does not get a contribution from the local operator. Equivalently, identifying $\sqrt{3}/2$ and $m^{2}$ with $\sin \theta_{W}$ and $M_{W}^{2}$ respectively and using (\ref{Sdef}) and (\ref{Tdef}), we expect that $S - 4\cos \theta_{W} \ T$ ($S - 2T$ in our model) is finite even in more than five dimensions.   

Let us confirm that this expectation really holds for 6D space-time. 
For such purpose, we focus on the (possibly) divergent parts of S and T parameters. 
For T-parameter, it is given by (\ref{divT}), 
\bea
T_{({\rm div})} &=& 
- \pi \frac{2^{\frac{3}{2}D-3}}{(4\pi)^{D/2}}
\frac{(1-2^{3-D})(D-1)}{D(3-D)}
\frac{\Gamma(\frac{5-D}{2})\Gamma(\frac{D-1}{2})^2}
{\Gamma(\frac{3}{2})\Gamma(D-1)}(2\pi R)m^{D-3}. 
\label{divTD}
\eea
As for S-parameter, it is given by (\ref{divS}), 
\bea
S_{({\rm div})}  
&=& - \frac{9\pi 2^{3D/2-5}}{(4\pi)^{D/2}\Gamma(5/2)}
\frac{D-1}{3-D}\frac{\Gamma \left(\frac{5-D}{2} \right) 
\Gamma \left(\frac{D+1}{2} \right)^2}{\Gamma(D+1)}(2\pi R)m^{D-3}. 
\label{divSD}  
\eea
From these expressions, we can find that 
the ratio indicated by the operator analysis (\ref{ratio}) 
indeed appears in 6D space-time, as we expected:  
\bea
S_{({\rm div})} = 
\frac{3(5-1)}{8(1-2^{3-5})} T_{({\rm div})} 
= 2 T_{({\rm div})}. 
\eea
Thus we have confirmed  $S - 2T$ is finite as we expected. 

We can also show that $S - 2T$ is finite in 6D case by using the results 
due to the momentum integration (by use of dimensional regularization) 
before the mode summation. 
Going back to the result (\ref{TKK2}), for the case of $D=5$ (6D), 
and Taylor expanding the integrand in the powers of $m/m_{n}$ 
up to ${\cal O}((m/m_n)^6)$, $T(6D)$ can be calculated as
\bea
T_{(n \neq 0)} (6D) &=& 
- \frac{\sqrt{2}}{5\pi}
\sum_{n=1}^\infty 
\left[
-\frac{m^2}{m_n} +\frac{1}{12}\frac{m^4}{m_n^3}
\right]. 
\label{6DT}
\eea
The first term is actually ``logarithmically" divergent, 
once K-K mode sum is taken. 
Similarly, $S(6D)$ is calculated from (\ref{Sdr}) 
up to ${\cal O}((m/m_n)^4)$,  
\bea
S_{(n \neq 0)}(6D)  
= - \frac{2\sqrt{2}}{5\pi} 
\sum_{n=1}^\infty 
\left[ -\frac{m^2}{m_n} +\frac{3}{14}\frac{m^4}{m_n^3}
\right].  
\label{6DS}
\eea
The first term is also ``logarithmically" divergent. 
By taking the specific linear combination of these results 
(\ref{6DT}) and (\ref{6DS}), 
we obtain a finite result (at the leading order),  
\bea
S_{(n \neq 0)}(6D) - 2 T_{(n \neq 0)}(6D) 
= 
\frac{11\sqrt{2}}{210\pi} 
m^4 R^3 \zeta(3)
\label{6Dpdtn}
\eea
where $\zeta(3) = \sum_{n=1}^\infty 1/n^3
= 1.2020569303 \cdots$. 

Some comments are in order. 
In our model on $M^{D} \times (S^{1}/Z_{2})$, only one extra spatial 
dimension is regarded to be compactified. 
One might think that our arguments of finiteness 
for higher than five dimensional cases 
($D > 4$) is meaningless, since the non-compact space-time is five dimensions 
not four dimensions for 6D case, for example. 
However, our argument with respect to the UV divergence will not be affected,  
irrespectively of the compactness of extra dimensions.  
This is because the information of compactification is a global aspect, 
namely the IR nature of the theory, so the structure of UV divergence 
has nothing to do with that. 
Therefore, the finiteness of the quantity 
$S - 4 \cos \theta_{W} \ T$ holds true even in the 6D theory compactified on $T^2/Z_2$, for example, 
although the remaining finite value itself might be changed. 

Another issue to be addressed is that the finiteness of 
$S - 4 \cos \theta_{W} \ T$ does not seem to hold for 
higher than six dimensional cases ($D > 5$), 
as suggested from (\ref{divTD}) and (\ref{divSD}). 
Let us note that, for more than six dimensional cases, 
each of S and T parameters gets divergent contributions 
also from the gauge invariant operators, 
whose mass dimensions are higher than six (from 4D point of view). 
Thus the divergent contributions come from the multiple operators 
and it is no longer possible to find out a finite observable 
in a model independent way.    

One-loop contributions to the S and T parameters also have been calculated 
in the UED scenario \cite{UED}, 
where these parameters become finite in five dimensions, 
but divergent in more than five dimensions. 
Thus, the gauge-Higgs unification and the UED scenarios 
share the same divergence structure at this point. 
However, as was shown above, 
the divergences of S and T parameters are not independent and 
a particular linear combination of these parameters is predictable 
in the gauge-Higgs unification, in a model independent way. 
On the other hand, in the UED scenario, 
even if some combination of the S and T parameters in 6D case is related, 
the combination will dependent on the detail (the choice of matter fields, etc.) 
of each model.   
This is essentially because in the UED scenario the operators responsible 
for the parameters are mutually independent as in the SM. 
This is the crucial difference between the gauge-Higgs unification scenario 
and the UED scenario.

\section{Summary and concluding remarks} 
In this paper, we have discussed the one-loop contributions to the 
S and T parameters in the gauge-Higgs unification scenario.  
Taking a minimal $SU(3)$ gauge-Higgs unification model 
with a triplet fermion as the matter fields, we have calculated  
the S and T parameters in two different approaches. 
One is an approach to perform the momentum integral 
by use of dimensional regularization before taking the K-K mode sum. 
The other approach is to take the mode sum first before the momentum integration. 
The former has a natural approach from the point of view 
to make the 4D local gauge symmetry and the custodial symmetry manifest. 
On the other hand, 
the latter approach also has an advantage to make 
the higher dimensional gauge invariance and the structure of UV divergences manifest. 

In five dimensional space-time, we have shown that the one-loop contributions to 
the S and T parameters are both finite, 
and evaluated their finite values explicitly, 
adopting two different approaches stated above. 
In more than five dimensions, we find that 
the S and T parameters themselves are divergent 
as in the Universal Extra Dimension (UED) scenario. 
However, we have derived a genuine prediction of the gauge-Higgs unification scenario, i.e. that a particular linear combination of the S and T parameters, $S - 4\cos \theta_{W} \ T$, 
is calculable (UV finite) for the case of six dimensional space-time. 
The relative ratio of the coefficients appearing in the linear combination 
turns out to coincide with what is derived from an analysis 
of single higher dimensional gauge invariant operator, and 
therefore is determined in a model independent way. 
This is the crucial difference from the situation in the UED scenario. 

The investigation done in this paper proves 
the predictability of the gauge-Higgs unification scenario 
concerning the S and T parameters, 
even though higher dimensional gauge theories are understood 
to be non-renormalizable. 
Thus, in order to verify the feasibility of the scenario 
and/or to search for the genuine predictions of the scenario, 
it is very interesting to study these parameters 
in more realistic gauge-Higgs unification models, 
having reasonable Weinberg angle and quark masses, 
and to extract the phenomenological consequences 
utilizing existing very precise data on the oblique parameters.  

It will be natural to expect that calculable observables controlled 
by the higher dimensional gauge invariance, other than the oblique parameters, 
still remain to be found in the scenario. 
We will continue to search for such observables.   

%
%
%
\subsection*{Acknowledgment}

The work of the authors was supported 
in part by the Grant-in-Aid for Scientific Research 
of the Ministry of Education, Science and Culture, No.18204024.  

\begin{appendix}

\setcounter{equation}{0}
\section{Derivation of the super-convergent part of T-parameter (\ref{5DscT})}
\label{scT}
In this appendix, we show the detailed calculations to 
arrive at the result (\ref{5DscT}). 
The starting point is 
\bea
T_{({\rm sc})} &=& 
\frac{\pi}{\alpha^{2}}\frac{2^{D/2}}{D}L^{4-D} 
\int_0^1dt \int\frac{d^D\rho}{(2\pi)^D} \nonumber \\
&& \times \left[
-\frac{D}{\rho} \left( \frac{\sinh \rho}{\cosh \rho -1} - 1 \right) 
-\frac{D}{2\rho}\left( \frac{\sinh \rho}{\cosh \rho -\cos \alpha} - 1 \right) 
\right. \nonumber \\
&& \left. 
-\frac{D}{2} 
\frac{\left(1 + (D-2) \frac{\alpha^2}{\rho^2} \right)}
{\sqrt{\rho^2 + 4t(1-t)\alpha^2}}
\left(
\frac{\sinh \sqrt{\rho^2 + 4t(1-t)\alpha^2}}
{\cosh \sqrt{\rho^2 + 4t(1-t)\alpha^2}-\cos[(2t-1)\alpha]}  -1 \right)
\right. \nonumber \\
&& \left. 
+\frac{D}{2}
\frac{\left(4 + (D-2) \frac{\alpha^2}{\rho^2} \right)}
{\sqrt{\rho^2 + t(1-t)\alpha^2}}
\left(
\frac{\sinh \sqrt{\rho^2 + t(1-t)\alpha^2}}
{\cosh \sqrt{\rho^2 + t(1-t)\alpha^2}-\cos [t \alpha]} -1 
\right) 
\right]. 
\eea
Using a formula 
\bea
\frac{1-x^2}{1 - 2x \cos \theta +x^2} = 
1+\sum_{n=1}^\infty 2(\cos n \theta) x^n, 
\label{f1}
\eea
we obtain
\bea
\int\frac{d^D\rho}{(2\pi)^D}\frac{1}{\rho}
\left(
\frac{\sinh \rho}{\cosh \rho -\cos \alpha}-1
\right)
&=& \frac{4\pi^{D/2}}{(2\pi )^D \Gamma(D/2)} \sum_{n=1}^\infty 
\frac{\cos(n\alpha)}{n^{D-1}}\int_0^\infty d\rho \rho^{D-2}e^{-\rho} 
\nonumber \\
&=& \frac{4}{(4\pi)^{D/2}}\frac{\Gamma(D-1)}{\Gamma(D/2)}
\sum_{n=1}^\infty \frac{\cos(n\alpha)}{n^{D-1}}. 
\eea
Next we consider the following type of integral, 
\bea
F(a, b, \theta) &\equiv& i\int\frac{d^D\rho}{(2\pi L)^D}\rho^{-a}
\frac{1}{\sqrt{\rho^2+b^2}}
\left(
\frac{\sinh\sqrt{\rho^2 + b^2}}{\cosh\sqrt{\rho^2 + b^2} - \cos \theta} - 1
\right), \\
&=& \frac{4i\pi^{D/2}}{(2\pi L)^D\Gamma(D/2)}\sum_{n=1}^\infty 
\cos(n\theta) \int_0^\infty d\rho \rho^{D-1-a}\frac{1}{\sqrt{\rho^2+b^2}} 
e^{-n\sqrt{\rho^2+b^2}}
\eea
where (\ref{f1}) is used in the second line. 
Rescaling $n \rho \to \rho$ and the change of the integration variable 
$\rho \to x=\sqrt{\rho^2+(nb)^2}$ lead to
\bea
F(a, b, \theta) = \frac{4iL^{-D}}{(4\pi)^D\Gamma(D/2)} 
\sum_{n=1}^\infty \frac{\cos(n\theta)}{n^{D-a-1}} 
\int_{nb}^\infty dx (x^2-(nb)^2)^{(D-2-a)/2}e^{-x}. 
\eea
For $D=4$ and $a=0$ or $2$, the above integral can be performed to get  
\bea
F(0,b,\theta) &=& \frac{8iL^{-4}}{(4\pi)^2} \sum_{n=1}^\infty 
\frac{\cos(n\theta)}{n^3}(nb+1)e^{-nb}, \\
&=& \frac{8iL^{-4}}{(4\pi)^2}
\left[
\zeta(3) +\frac{\theta^2+b^2}{4}\ln(\theta^2 + b^2) 
-\frac{3}{4}\theta^2 -\frac{1}{4}b^2 -\frac{1}{6}b^3 -\frac{1}{288}\theta^4 
\right. \nonumber \\
&& \left. 
-\frac{1}{48}\theta^2 b^2 +\frac{1}{96}b^4 + \cdots
\right], 
\label{exp1} \\
F(2,b,\theta) &=& \frac{4iL^{-4}}{(4\pi)^2} \sum_{n=1}^\infty 
\frac{\cos(n\theta)}{n}e^{-nb}, \\
&=& \frac{4i L^{-4}}{(4\pi)^2} 
\left[
-\frac{1}{2}\ln(b^2 + \theta^2) +\frac{1}{2}b - \frac{1}{24}b^2 
+ \frac{1}{24} \theta^2 + \cdots
\right], 
\label{exp2}
\eea
where the following expansion formula for small $\theta$ and $b$ 
are used in the second line. 
\bea
\sum_{n=1}^\infty \frac{\cos(n \theta)}{n^2}e^{-nb} 
&=& \frac{\pi^2}{6} +\frac{b}{2}\ln(\theta^2+b^2) 
+ \theta \tan^{-1} \left(\frac{b}{\theta} \right) 
- \frac{\pi}{2}\theta -b \nonumber \\
&& + \frac{\theta^2}{4} -\frac{b^2}{4} -\frac{\theta^2 b}{24} 
+ \frac{b^3}{72} + \cdots, \\
\sum_{n=1}^\infty \frac{\cos(n \theta)}{n^3} e^{-nb} 
&=& 
\zeta(3) -\frac{\pi^2}{6}b + \frac{\theta^2 - b^2}{4} \ln(\theta^2 + b^2) 
-\theta b \tan^{-1} \left( \frac{b}{\theta} \right) 
-\frac{3}{4}(\theta^2 - b^2) \nonumber \\
&& + \frac{\pi}{2} \theta b 
- \frac{1}{4} \theta^2 b 
+ \frac{b^3}{12} - \frac{\theta^4}{288} 
+ \frac{\theta^2 b^2}{48} 
-\frac{b^4}{288} + \cdots. 
\eea
Thus, $T_{({\rm sc})}$ for 5D case is given by
\bea
T_{({\rm sc})}~(5D) &=& 
\frac{1}{4\pi \alpha^{2}} \int_0^1 dt 
\left[
-8\zeta(3) -4 
\left(
\zeta(3) +\frac{\alpha^2}{4}\ln \alpha^2 -\frac{3}{4}\alpha^2 
-\frac{\alpha^4}{288}
\right) \right. \nonumber \\
&& \left. 
-2\tilde{F}(0, \sqrt{4t(1-t)}\alpha, (2t-1)\alpha) 
-4\alpha^2 \tilde{F}(2, \sqrt{4t(1-t)}\alpha, (2t-1)\alpha) 
\right. \nonumber \\
&& \left. +8\tilde{F}(0, \sqrt{t(1-t)}\alpha, t\alpha) 
+4\alpha^2 \tilde{F}(2, \sqrt{t(1-t)}\alpha, t\alpha) 
\right]
\eea
where 
\bea
F(a, b, \theta) &\equiv& \frac{4iL^{-4}}{(4\pi)^2}\tilde{F}(a,b,\theta). 
\eea
From (\ref{exp1}) and (\ref{exp2}), we find 
\bea
T_{({\rm sc})} (5D) &\simeq& \frac{1}{4\pi \alpha^{2}} \int_0^1 dt 
\left[
(-12t + 6 + 4t \ln t -2\ln t)\alpha^2 \right. \nonumber \\
&& \left. 
+ \left(\frac{8}{3}(t(1-t))^{3/2} -2\sqrt{t(1-t)} \right) \alpha^3 
\right. \nonumber \\
&& \left. + \frac{1}{36}(-48t^4 + 104 t^3 - 102t^2 + 50t - 5)\alpha^4 
\right], \\
&=& \frac{1}{4\pi m^{2}} 
\left(m^2 -\frac{3\pi^{2}}{8}(mR)m^2 + \frac{4\pi^2}{15}(mR)^2m^2 \right). 
\eea

\end{appendix}

\end{document}